\documentclass[10pt]{article}

\usepackage{graphicx,color}
\begin{document}

\title{Search in Power-Law Networks}
\author{Lada A. Adamic 
\thanks{email: ladamic@stanford.edu}
\\ Stanford University
\\ Dept. of Applied Physics
\and 
    Rajan M. Lukose
    \thanks{email: lukose@hpl.hp.com}
\\ HP Sand Hill Labs
\\ Palo Alto, CA 94304
\and
    Amit R. Puniyani
    \thanks{email: amit8@stanford.edu}
    \\ Stanford University
\\ Dept. of Physics
\and 
    Bernardo A. Huberman
    \thanks{email: huberman@hpl.hp.com}
\\ HP Sand Hill Labs
\\ Palo Alto, CA 94304
} 

\maketitle

\begin{abstract}
Many communication and social networks have power-law link
distributions, containing a few nodes which have a very high
degree and many with low degree. The high connectivity nodes play
the important role of hubs in communication and networking, a fact
which can be exploited when designing efficient search algorithms.
We introduce a number of local search strategies which utilize
high degree nodes in power-law graphs and which have costs which
scale sub-linearly with the size of the graph. We also demonstrate
the utility of these strategies on the Gnutella peer-to-peer
network.
\end{abstract}

\section{Introduction}

A number of large distributed systems, ranging from social
\cite{aiello00massrandom} to communication
\cite{faloutsos99topology} to biological networks
\cite{barabasi00metabolic} display a power-law distribution in
their node degree. This distribution reflects the existence of a
few nodes with very high degree and many with low degree, a
feature not found in standard random graphs
\cite{erdos60randgraph}. An illustration of the power-law nature
of such networks is given by the AT\&T call graph. A call graph is
a graph representation of telephone traffic on a given day in
which nodes represent people and links the phone calls among them.

\begin{figure}[fanOutDegree]
\begin{center}
\includegraphics[scale=0.15]{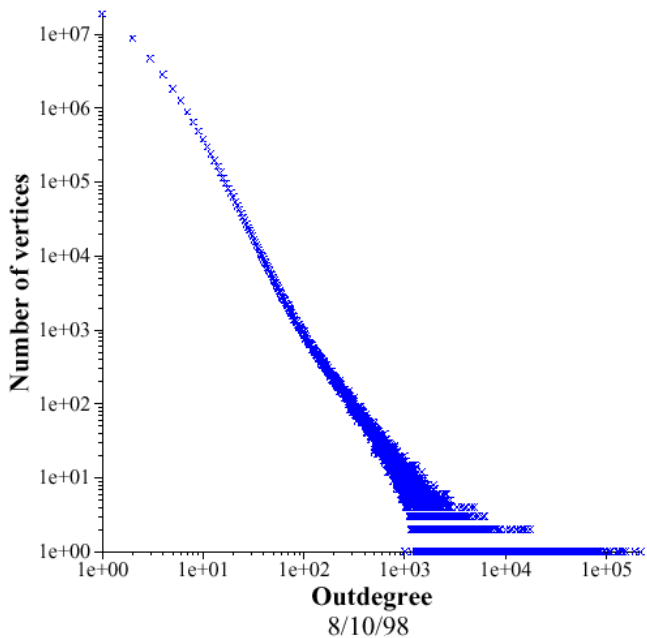}
\end{center}
\caption[]{Out-degree distribution for the AT\&T call graph.}
\label{fanOut}
\end{figure}

As is shown in Figure \ref{fanOut}, the out-link degree
distribution for a massive graph of telephone calls between
individuals is power-law, with an exponent of approximately 2.1.
The same distribution is obtained for the case of in-links. This
power-law in the link distribution reflects the presence of
central individuals who interact with many others on a daily basis
and play a key role in relaying information.

While recent work has concentrated on the properties of these 
power-law networks and how they are dynamically generated 
\cite{barabasi99scaling,adamic99nature,newman01graphs}, there 
remains the interesting problem of finding efficient algorithms 
for searching within these particular kinds of graphs. Recently, 
Kleinberg \cite{kleinberg00nav} studied search algorithms in a 
graph where nodes are placed on a 2-D lattice and each node has a 
fixed number of links whose placement is correlated with lattice 
distance to the other nodes. Under a specific form of the 
correlation, an algorithm with knowledge of the target's location 
can find the target in polylogarithmic time. 

In the most general distributed search context however, one may
have very little information about the location of the target.
Increasingly a number of pervasive electronic networks, both wired
and wireless, make geographic location less relevant.  A
particularly interesting example is provided by the recent
emergence of peer-to-peer networks, which have gained enormous
popularity with users wanting to share their computer files. In
such networks, the name of the target file may be known, but due
to the network's ad hoc nature, until a real-time search is
performed the node holding the file is not known. In contrast to
the scenario considered by Kleinberg, there is no global
information about the position of the target, and hence it is not
possible to determine whether a step is a move towards or away
from the target. One simple way to locate files, implemented by
Napster, is to use a central server which contains an index of all
the files every node is sharing as they join the network. This is
the equivalent of having a giant white pages for the entire United
States. Such directories now exist online, and have in a sense
reduced the need to find people by passing messages. But for
various reasons, including privacy and copyright issues, in a
peer-to-peer network it is not always desirable to have a central
server.

File-sharing systems which do not have a central server include
Gnutella and Freenet. Files are found by forwarding queries to
one's neighbors until the target is found. Recent measurements of
Gnutella networks \cite{clip200bwbarrier} and simulated Freenet
networks \cite{hong01freenet} show that they have power-law degree
distributions. In this paper, we propose a number of
message-passing algorithms that can be efficiently used to search
through power law networks such as Gnutella. Like the networks
that they are designed for, these algorithms are completely
decentralized and exploit the power law link distribution in the
node degree. The algorithms use local information such as the
identities and connectedness of their neighbors, and their
neighbors' neighbors, but not the target's global position. We
demonstrate that our search algorithms work well on real Gnutella
networks, scale sub-linearly with the number of nodes, and may
help reduce the network search traffic that tends to cripple such
networks.

The paper is organized as follows.  In Section 2, we present
analytical results on message-passing in power-law graphs,
followed by simulation results in Section 3.  Section 4 compares
the results with Poisson random graphs.  In Section 5 we consider
the application of our algorithms to Gnutella, and Section 6
concludes.

\section{Search in Power-Law Random Graphs}
In this section we use the generating function formalism
introduced by Newman \cite{newman01graphs} for graphs with
arbitrary degree distributions to analytically characterize
serach-cost scaling in power-law graphs.

\subsection{Random Walk Search}
Let $G_0(x)$ be the generating function for the distribution of
the vertex degrees $k$. Then
\begin{equation}
G_0(x) = \sum_{0}^{\infty}p_k x^k \label{G0defined}
\end{equation}
where $p_k$ is the probability that a randomly chosen vertex on
the graph has degree $k$.

For a graph with a power-law distribution with exponent $\tau$,
minimum degree $k = 1$ and an abrupt cutoff at $m = k_{max}$, the
generating function is given by

\begin{equation}
G_0(x) = c \sum_{1}^{m}k^{-\tau} x^k   \label{G0power}
\end{equation}

with $c$ a normalization constant which depends on $m$ and $\tau$
to satisfy the normalization requirement

\begin{equation}
G_0(1) = c \sum_{1}^{m} k^{-\tau} = 1
\end{equation}

The average degree of a randomly chosen vertex is given by

\begin{equation}
<k> = \sum_{1}^{m} k p_{k} = G_0^{\prime}(1)
\end{equation}

Note that the average degree of a vertex chosen at random and one
arrived at by following a random edge are different. A random edge
arrives at a vertex with probability proportional to the degree of
the vertex, i.e. $p^\prime(k) \sim k p_k$. The correctly
normalized distribution is given by

\begin{equation}
\frac{\sum_k k p_k x^k}{\sum_k k p_k} = x \frac{G^\prime_0
(x)}{G^\prime_0 (1) }
\end{equation}

If we want to consider the number of outgoing edges from the
vertex we arrived at, but not include the edge we just came on, we
need to divide by one power of $x$. Hence the number of new
neighbors encountered on each step of a random walk is given by
the generating function

\begin{equation}
G_1(x) = \frac{G^\prime_0(x)}{G^\prime_0 (1)}
\end{equation}

where $G^\prime_0 (1)$ is the average degree of a randomly chosen
vertex as mentioned previously.

In real social networks it is reasonable that one one would have
at least some knowledge of one's friends' friends. Hence, we now
compute the distribution of second degree neighbors. The
probability that any of the 2nd degree neighbors connect to any of
the first degree neighbors or to one another goes as $N^{-1}$ and
can be ignored in the limit of large $N$. Therefore, the
distribution of the second neighbors of the original randomly
chosen vertex is determined by

\begin{equation}
\sum_k p_k[G_1(x)]^k = G_0(G_1(x))
\end{equation}

It follows that the average number of second degree neighbors is
given by

\begin{equation}
z_{2A} = [\frac{\partial}{\partial x} G_0(G_1(x))]_{x=1} =
G^\prime_0 (1) G^\prime_1 (1) \label{2ndDegNeighRand}
\end{equation}

Similarly, if the original vertex was not chosen at random, but
arrived at by following a random edge, then the number of second
degree neighbors would be given by

\begin{equation}
z_{2B} = [\frac{\partial}{\partial x} G_1(G_1(x))]_{x=1} =
[G^\prime_1 (1)]^2 \label{2ndDegNeighFollow}
\end{equation}

In both Equation \ref{2ndDegNeighRand} and Equation
\ref{2ndDegNeighFollow} the fact that $G_1 (1) = 1$ was used.

Both these expressions depend on the values $G^\prime_0 (1)$ and
$G^\prime_1 (1)$ so let's calculate those for given $\tau$ and
$m$. For simplicity and relevance to most real-world networks of
interest we assume $2 < \tau < 3$.

\begin{equation}
G^\prime_0 (1) = \sum_1^m c k^{1-\tau} \sim \int_1^m x^{\tau -1}
dx = \frac{1}{\tau-2}(1 - m^{2-\tau})
\end{equation}

\begin{eqnarray}
G^\prime_1 (1) & = &\frac{1}{G^\prime_0 (1)} \frac{\partial}{\partial x} \sum_1^m c k^{1-\tau} x^{k-1} \\
\\ & = & \frac{1}{G^\prime_0 (1)}\sum_2^m c k^{1-\tau} (k-1) x^{k-2}
\\ & \sim & \frac{1}{G^\prime_0 (1)} (\frac{m^{3-\tau} (\tau - 2) -
2^{2-\tau} (\tau - 1) + m^{2 - \tau} (3 - \tau)}{(\tau-2)(3-\tau)}
\end{eqnarray}

for large cutoff values $m$. Now we impose the cutoff of Aiello et
al. \cite{aiello00massrandom} at $m \sim N^{1/\tau}$. Since $m$
scales with the size of the graph $N$ and for $2 < \tau < 3$ the
exponent $2-\tau$ is negative, we can neglect terms constant in
$m$. This leaves

\begin{equation}
G^\prime_1 (1) \\ = \frac{1}{G^\prime_0 (1)}
\frac{m^{3-\tau}}{(3-\tau)}
\end{equation}

Substituting into Equation \ref{2ndDegNeighRand} (the starting
node is chosen at random) we obtain
\begin{equation}
z_{2A} = G^\prime_0 (1) G^\prime_1 (1) \sim m^{3-\tau}
\end{equation}

We can also derive $z_{2B}$, the number of 2nd degree neighbors
encountered as one is doing a random walk on the graph.

\begin{equation}
z_{2B} = [G^\prime_1 (1)]^2 = [\frac{\tau-2}{1-m^{2-\tau}}
\frac{m^{3-\tau}}{3-\tau}]^2 \label{G12eq}
\end{equation}

Letting $m \sim N^{1/\tau}$ as above, we obtain

\begin{equation}
z_{2B} \sim N^{2 (\frac{3}{\tau} - 1)}
\end{equation}

and the scaling of the number of steps required becomes

\begin{equation}
s \sim N^{3(1-2/\tau)} \label{RandStratScale}
\end{equation}

In the limit $\tau \rightarrow 2$, equation \ref{G12eq} becomes

\begin{equation}
z_{2B} \sim \frac{N}{ln^2(N)}
\end{equation}

and the scaling of the number of steps required is

\begin{equation}
s \sim ln^2(N)
\end{equation}

\subsection{Search utilizing high degree nodes}

Random walks in power-law networks naturally gravitate towards the
high degree nodes, but an even better scaling is achieved by
intentionally choosing high degree nodes. For $\tau$ sufficiently
close to $2$ one can walk down the degree sequence, visiting the
node with the highest degree, followed by a node of the next
highest degree, etc. Let $m - a$ be the degree of the last node we
need to visit in order to scan a certain fraction of the graph. We
make the self-consistent assumption that $a << $m, i.e. the 
degree of the node hasn't dropped too much by the time we have 
scanned a fraction of the graph. Then the number of first degree 
neighbors scanned is given by

\begin{equation}
z_{1D} = \int^{m}_{m-a} N k^{1-\tau} dk \sim N a m^{1-\tau}
\end{equation}

The number of nodes having degree between $m-a$ and $m$, or
equivalently, the number of steps taken is given by
$\int^{m}_{m-a} k^{-\tau} \sim a$. The number of second degree
neighbors when one follows the degree sequence is given by:

\begin{equation}
z_{1D} \ast G^{\prime}_{1}(1) \sim N a m^{2(2-\tau)}
\end{equation}

which gives the number of steps required as

\begin{equation}
s \sim m^{2(\tau-2)} \sim N^{2-\frac{4}{\tau}}
\label{DegStratScale}
\end{equation}

We now consider when and why it is possible to go down the degree
sequence. We start with the fact that the original degree
distribution is a power-law:
\begin{equation}
p(x) = (\sum_{1}^{m} x^{-\tau})^{-1} x^{-\tau}
\end{equation}

where $m=N^{1/\tau}$ is the maximum degree. A node chosen by
following a random link in the graph will have its remaining
outgoing edges distributed according to

\begin{equation}
p^\prime(x) = (\sum_{0}^{m-1}
(x+1)^{(1-\tau)})^{-1}(x+1)^{(1-\tau)}
\end{equation}

At each step one can choose the highest degree node among the $n$
neighbors. The expected number of the outgoing edges of that node
can be computed as follows. In general, the cumulative
distribution (CDF) \(P_{max}(x,n) \) of the maximum of n random
variables can be expressed in terms of the CDF \( P(x) = \int_0^x
p(x\prime)\, \mathrm{d}x\prime \) of those random variables:
\(P_{max}(x,n)=P(x)^n \). This yields
\begin{equation}
p^\prime_{max}(x,n) = n (1+x)^{1-\tau} (\tau - 2) (1 - (x+1)^{2 -
\tau})^{n-1} (1 - N^{2/\tau - 1})^{-n} \label{PMaxDist}
\end{equation}
for the distribution of the number of links the richest neighbor
among n neighbors has.

Finally, the expected degree of the richest node among n is given
by
\begin{equation}
E[x_{max}(n)]=\sum^{m-1}_0 x p^\prime_{max}(x,n)
\end{equation}

\begin{figure}[tbp]
\begin{center}
\includegraphics[scale=1.25]{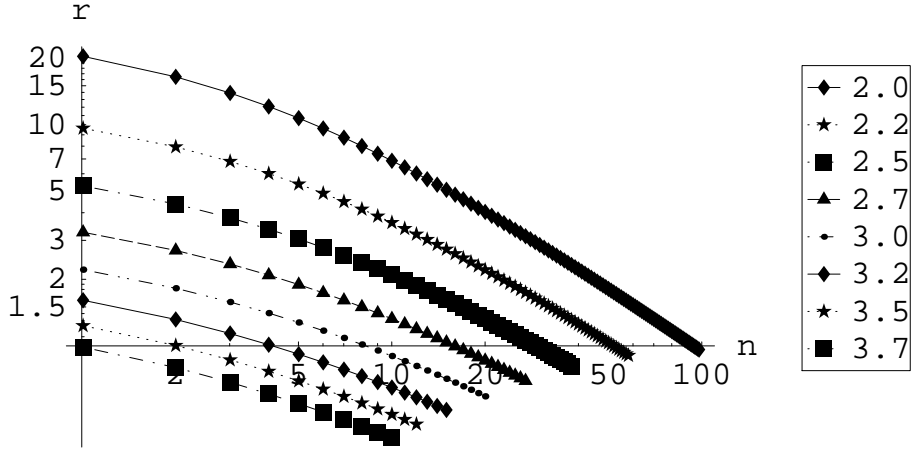}
\end{center}
\caption[Ratio of richest neighbor and the node itself for various
power-law exponents]{Ratio $r$ (the expected degree of the richest
neighbor of a node whose degree is $n$ divided by $n$) vs. $n$ for
$\tau$ (top to bottom) $=$ 2.0, 2.25, 2.5, 2.75, 3.00, 3.25, 3.50,
and 3.75. Each curve extends to the cutoff imposed for a 10,000
node graph with the particular exponent.} \label{degreeratios}
\end{figure}

We numerically integrated the above equation to derive the ratio
between the degree of a node and the expected degree of its
richest neighbor. The ratio is plotted in Figure
\ref{degreeratios}. For a range of exponents and node degrees, the
expected degree of the richest neighbor is higher than the node
itself. However, eventually (the precise point depends strongly on
the power-law exponent), the probability of finding an even higher
degree node in a neighborhood of a very high degree node starts 
falling.

What this means is that one can approximately follow the degree
sequence across the entire graph for a sufficiently small graph or
one with a power-law exponent close to 2 ($ 2.0 < \tau < 2.3$).
At each step one chooses a node with degree higher than the
current node, quickly finding the highest degree node. Once the
highest degree node has been visited, it will be avoided, and a
node of approximately second highest degree will be chosen.
Effectively, after a short initial climb, one goes down the
degree sequence. This is the most efficient way to do this kind
of sequential search, visiting highest degree nodes in sequence.

\section{Simulation}
We used simulations of a random social network with a power-law
link distribution of $\tau = 2.1$ to validate our analytical
results. As in the analysis above, a simple cutoff at $m \sim
N^{1/\tau}$ was imposed. The expected number of nodes among N
having exactly the cutoff degree is 1. No nodes of degree higher
than the cutoff are added to the graph. In real world graphs one
of course does observe nodes of degree higher than this imposed
cutoff, so that our simulations become a worse case scenario. Once
the graph is generated, the largest connected component (LCC) is
extracted, that is the largest subset of nodes such that any node
can be reached from any other node. For $2 < \tau < 3.48$ a giant
connected component exists \cite{aiello00massrandom}, and all our
measurements are performed on the LCC. We observe that the LCC
contains the majority of the nodes of the original graph and most
of the links as well. The link distribution of the LCC is nearly
identical to that of the original graph with a slightly smaller
number of 1 and 2 degree nodes.

Next we apply our message passing algorithm to the network. Two
nodes, the source and the target, are selected at random. At each
time step the node which has the message passes it on to one of
its neighbors. The process ends when the message is passed on to a
neighbor of the target. Since each node knows the identity of all
of its neighbors, it can pass the message directly to the target
if the target happens to be one of it's neighbors. The process is
analogous to performing a random walk on a graph, where each node
is 'visited' as it receives the message.

There are several variants of the algorithm, depending on the
strategy and the amount of local information available.
\begin{enumerate}
\item The node can pass the message on to one of its neighbors at
random, or it can avoid passing it on to a node which has already
seen the message.

\item If the node knows how many neighbors it has, it can choose to
pass the message onto the neighbor with the most neighbors.

\item The node may know only its neighbors, or it may know who its neighbors'
neighbors are. In the latter case it would pass the message onto a
neighbor of the target.
\end{enumerate}

In order to avoid passing the message to a node that has already
seen the message, the message itself must be signed by the nodes
as they receive the message. Further, if a node has passed the
message, and finds that all of its neighbors are already on the
list, it puts a special mark next to its name, which means that it
is unable to pass the message onto any new node. This is
equivalent to marking nodes as follows:

\begin{description}
\item[white] Node hasn't been visited.
\item[gray] Node has been visited, but all its neighbors have not.
\item[black] Node and all its neighbors have been visited already.
\end{description}

Here we compare two strategies. The first performs a random walk,
where only retracing the last step is disallowed. In the message
passing scenario, this means that if Bob just received a message
from Jane, he wouldn't return the message to Jane if he could pass
it to someone else. The second strategy is a self avoiding walk
which prefers high degree nodes to low degree ones. In each case
both the first and second degree neighbors are scanned at each
step.

\begin{figure}[tbp]
\unitlength1cm
\begin{minipage}[t]{7.4cm}
\begin{center}
\includegraphics[scale=0.4]{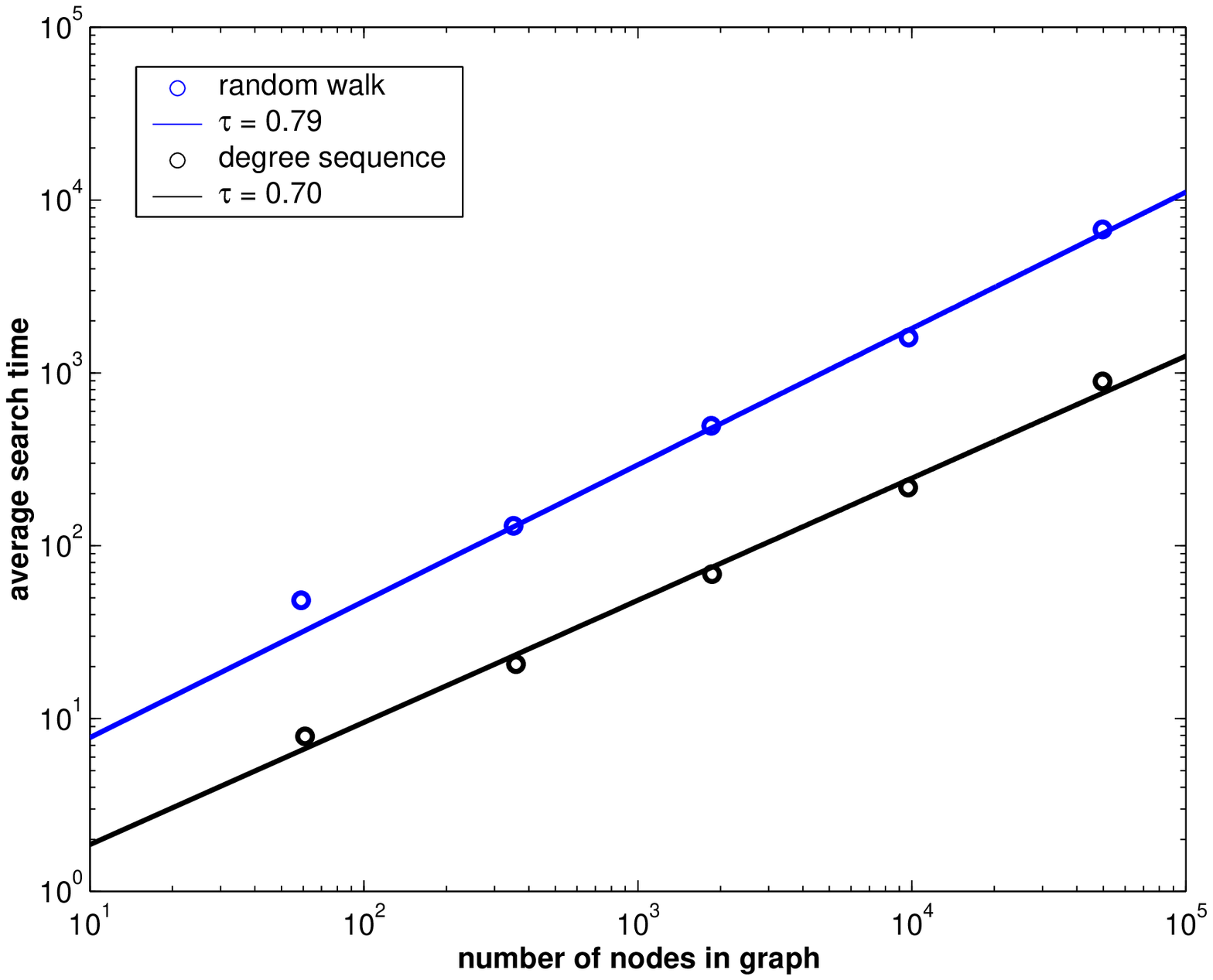}
\caption[Average node to node search time] {Scaling of the average
node to node search time in a random power-law graph with exponent
2.1, for random and high degree seeking strategies.
\label{AvSearchTime}}
\end{center}
\end{minipage}
\hfill
\begin{minipage}[t]{7.4cm}
\begin{center}
\includegraphics[scale=0.4]{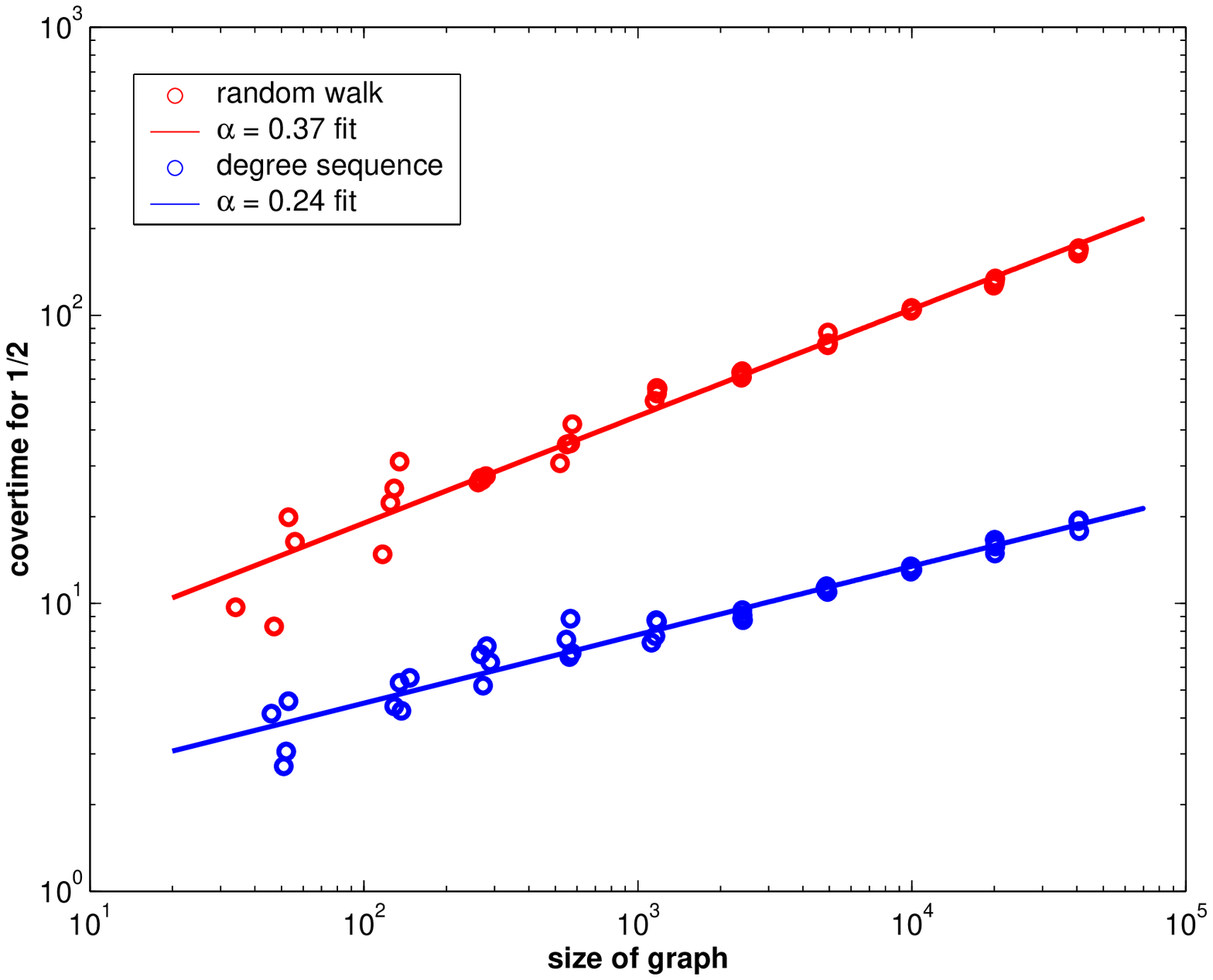}
\caption[Cover time for half of the graph]{Scaling of the time
required to cover one half the graph for random and high degree
seeking strategies. \label{HalfCoverTime}}
\end{center}
\end{minipage}
\end{figure}

Figure \ref{AvSearchTime} shows the scaling of the average search
time with the size of the graph for the two strategies. The
scaling (exponent 0.79 for the random walk and 0.70 for the high
degree strategy) is not as favorable as in the analytic results
derived above ($0.14$ for the random walk and 0.1 for the high
degree strategy when $\tau = 2.1$) .

Consider, on the other hand, the number of steps it takes to cover
half the graph.  For this measure  we observe a scaling which is
much closer to the ideal. As shown in Figure \ref{HalfCoverTime},
the cover time scales as $N^{0.37}$ for the random walk strategy
vs. $N^{0.15}$ from Equation \ref{RandStratScale}. Similarly, the
high degree strategy cover time scales as $N^{0.24}$ vs. $N^{0.1}$
in Equation \ref{DegStratScale}.

\begin{figure}[tbp]
\unitlength1cm
\begin{minipage}[t]{7.4cm}
\begin{center}
\includegraphics[scale=0.4]{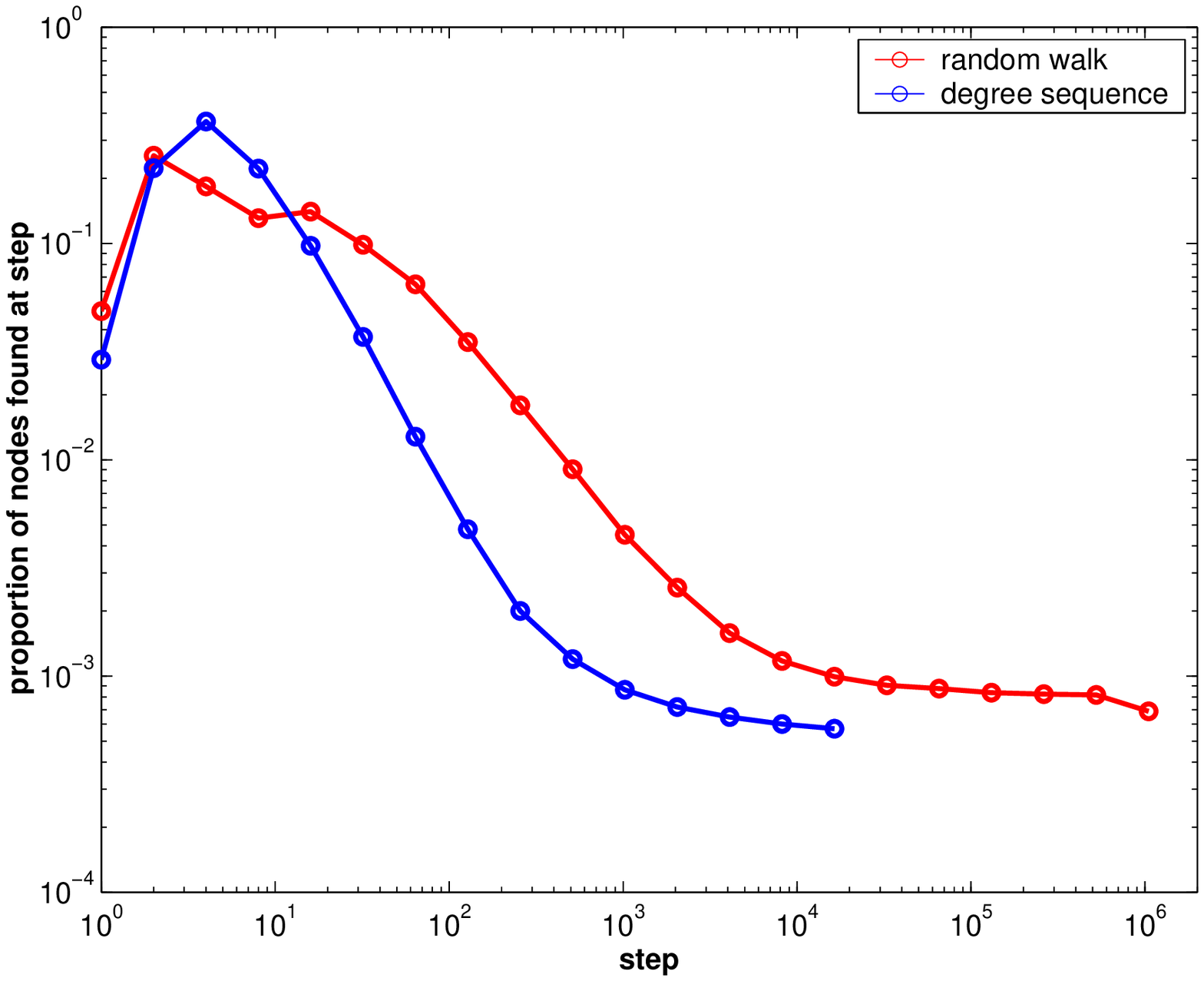}
\caption[Second neighbor search distribution]{Distribution of 
number of nodes seen as a function of the number of steps taken 
for random walk and high degree strategies for a 10,000 node 
graph. \label{StepDist} }
\end{center}
\end{minipage}
\hfill
\begin{minipage}[t]{7.4cm}
\begin{center}
\includegraphics[scale=0.4]{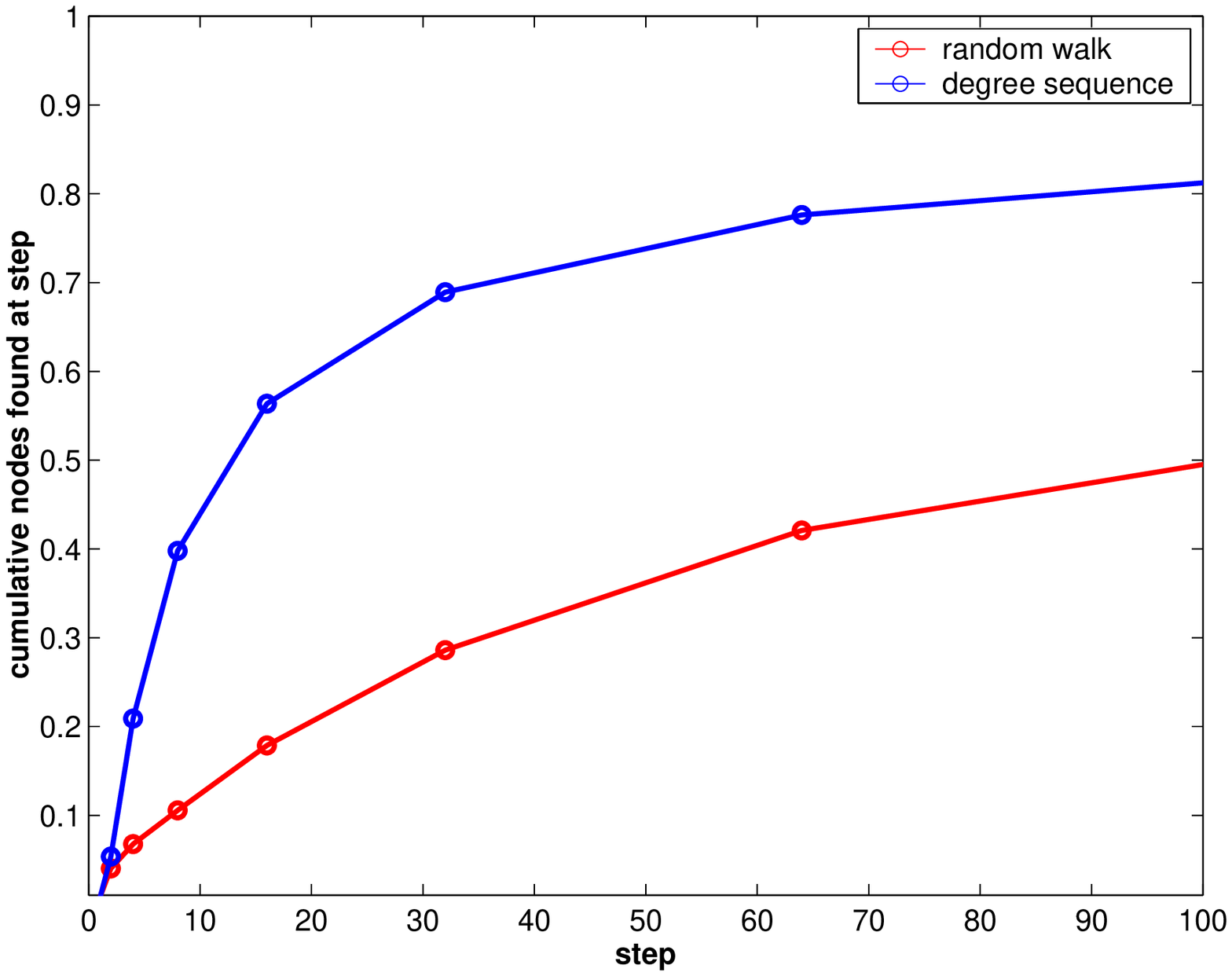}
\caption[Second neighbor cumulative search 
distribution]{Cumulative distribution of number of nodes seen as 
a function of the number of steps taken for random walk and high 
degree strategies for a 10,000 node graph. \label{CumStepDist}}
\end{center}
\end{minipage}
\end{figure}

The difference in the value of the scaling exponents of the cover
time and average search time implies that a majority of nodes can
be found very efficiently, but others demand high search costs.
Figures \ref{StepDist} and \ref{CumStepDist} show just that. A
large portion of the 10,000 node graph is covered within the first
few steps, but some nodes take as many steps or more to find as
there are nodes total. Take for example the high degree seeking
strategy. About 50\% of the nodes are scanned within the first 10
steps (meaning that it would take about  $10 + 2 = 12$ hops to
reach 50\% of the graph). However, the skewness of the search time
distribution bring the average number of steps needed to 217.

\begin{figure}[tbp]
\unitlength1cm
\begin{minipage}[t]{7.4cm}
\begin{center}
\includegraphics[scale=0.39]{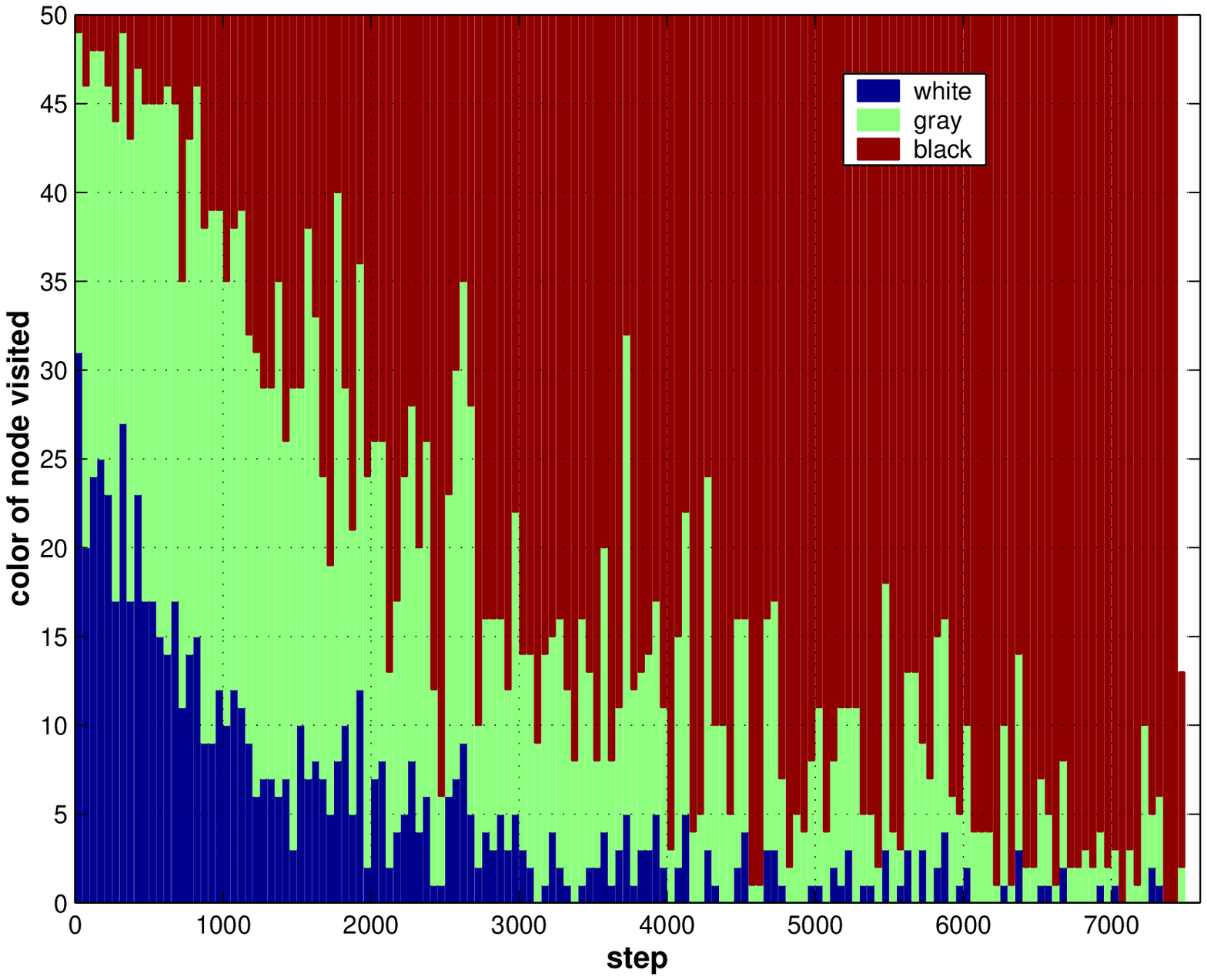}
\caption[Bar graph of the color of nodes visited in a random walk]
{Bar graph of the color of the nodes visited in a random walk on a
1,000 node power-law graph with exponent 2.1.
\label{bargraphcolorr} }
\end{center}
\end{minipage}
\hfill
\begin{minipage}[t]{7.4cm}
\begin{center}
\includegraphics[scale=0.39]{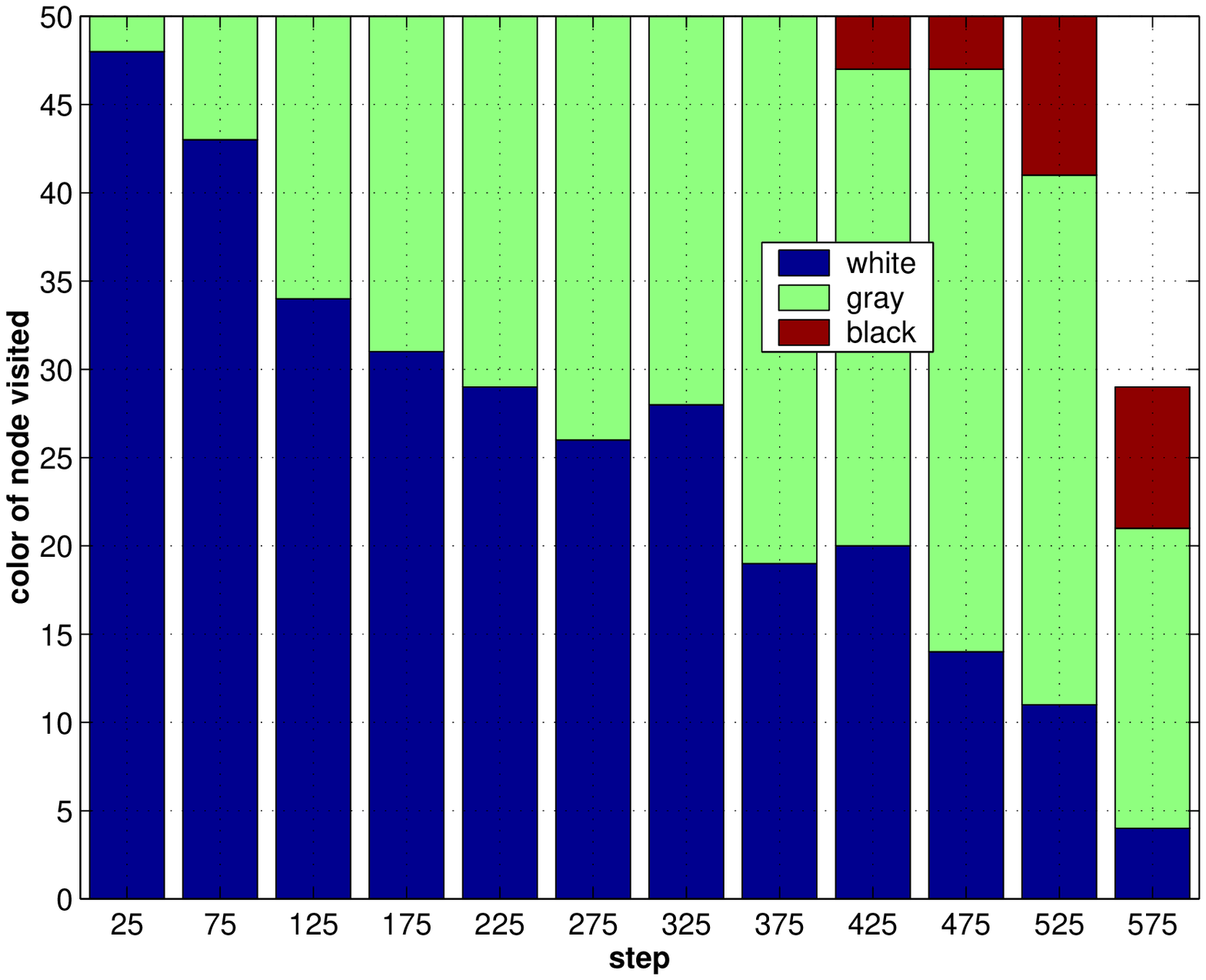}
\caption[Bar graph of the color of nodes visited in strategic
search] {Bar graph of the color of nodes visited in strategic
search of a random 1,000 node power-law graph with exponent 2.1.
\label{bargraphcolor}}
\end{center}
\end{minipage}
\end{figure}

Figure \ref{bargraphcolorr} illuminates why some nodes take such a
long time to find. It shows the color of nodes visited in a random
walk on a $N = 1000$ node power-law graph with exponent 2.1 and an
abrupt cutoff at $N^{1/2.1}$. The number of nodes of each color
encountered in 50 step segments is recorded in the bar for that
time period.

The brief initial period where many white nodes are visited is
followed by a long period dominated by gray and black nodes. The
random walk starts retracing its steps, only occasionally
encountering a new node it hasn't seen. Close to the end of the
walk, even the gray nodes have been exhausted as the random walk
repeatedly visits back nodes before it finds the final few nodes.

\begin{figure}[tbp]
\begin{center}
\includegraphics[scale=0.5]{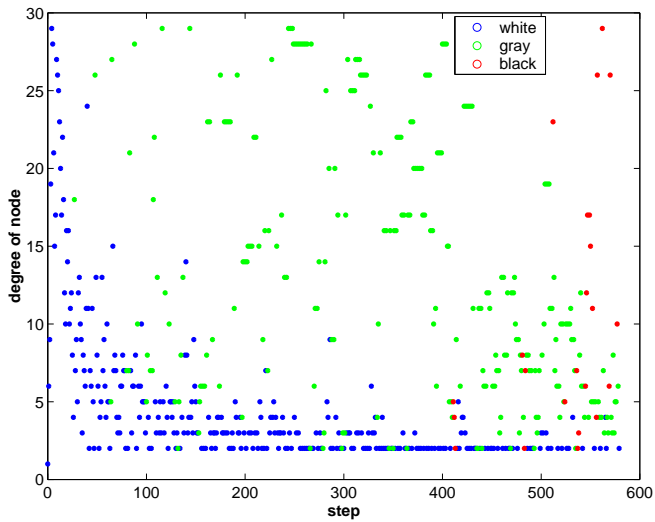}
\end{center}
\caption[Color and degree of nodes visited in strategic search]
{Color and degree of nodes visited in a strategic search of a
random 1,000 node power-law graph with exponent 2.1.}
\label{degreeandstep}
\end{figure}

A self-avoiding high-degree seeking random walk is an improvement
over the random walk described above, but still cannot avoid
retracing its steps. Figure \ref{bargraphcolor} shows the
corresponding bar chart in 50 step increments. Comparing figures
\ref{bargraphcolorr} and \ref{bargraphcolor} we observe that the
self-avoiding strategy is somewhat effective, with the total
number of steps needed to cover the graph about 13 times smaller,
and the fraction of visits to gray and black nodes significantly
reduced. Figure \ref{degreeandstep} shows the degree in addition
to the color of the node visited at every step. It demonstrates
that one is able to follow the degree sequence initially, going
from high degree white nodes to low degree white nodes.

Although the strategy of revisiting nodes modifies the scaling
behavior, it is not the process of revisiting itself which
necessarily changes the scaling. If nodes were uniformly linked,
at every step the number of new nodes seen would be proportional
to the number of unexplored nodes in the graph. The factor by
which the search is slowed down through revisits would be
independent of the size of the graph. Hence, revisiting alone does
not account for the difference in scaling.

The reason why the simulated scaling exponents for these search
algorithms do not follow the ideal is the same reason why
power-law graphs are so well suited to search: the link
distribution is extremely uneven. A large number of links point to
only a small subset of high degree nodes. When a new node is
visited, its links do not let us uniformly sample the graph, they
preferentially lead to high degree nodes, which have likely been
seen or visited in a previous step. This would not be true of a
Poisson graph, where all the links are randomly distributed and
hence all nodes have approximately the same degree. We will
explore and contrast the search algoritm on a Poisson graph in the
following section.

\section{Comparison with Poisson Distributed Graphs}
\begin{figure}[tbp]
\begin{center}
\includegraphics[scale=0.5]{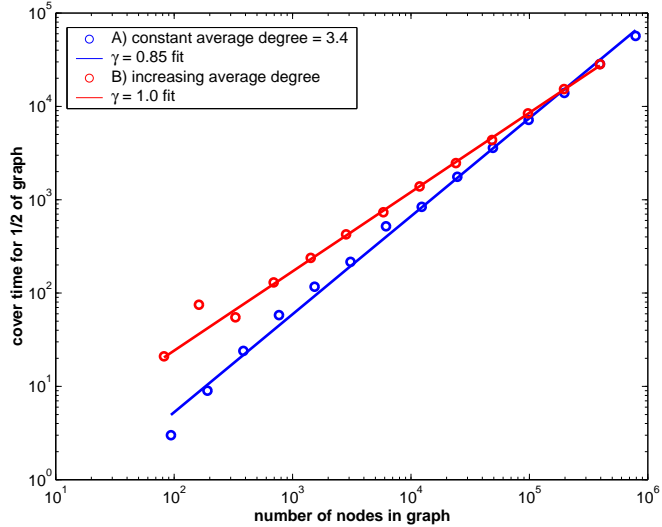}
\end{center}
\caption[Search time scaling in a Poisson graph]{Scaling of cover
time for 1/2 of the graph for a Poisson graph with A) a constant
average degree/node, and B) the same average degree/node as a
power-law graph with exponent 2.1.} \label{poissonscaling}
\end{figure}

In a Poisson random graph with $N$ nodes and $z$ edges, the
probability $p = z/N$ of an edge between any two nodes is the same
for all nodes. The generating function $G_0(x)$ is given by
\cite{newman01graphs}:
\begin{equation}
G_0(x) = e^{z(x-1)}
\end{equation}
In this special case $G_0(x) = G_1(x)$, so that the distribution
of outgoing edges of a node is the same whether one arrives at the
vertex by following a link or picks the node at random. This makes
analysis search in a Poisson random graph particularly simple. The
expected number of new links encountered at each step is a
constant $p$. So that the number of steps needed to cover a
fraction $c$ of the graph is $s = c N/p$. If $p$ remains constant
as the size of the graph increases, the cover time scales linearly
with the size of the graph. This has been verified via simulation
as shown in Figure \ref{poissonscaling}.

In our simulations the probability $p$ grows slowly towards its
asymptotic value as the size of the graph is increased because of
the particular choice of cutoff at $m \sim N^{(1/\tau)}$ for the
power-law link distribution. We generated Poisson graphs with the
same number of nodes and links for comparison. Within this range
of graph sizes, growth in the average number of links per node
appears as $N^{0.6}$, making the average number of 2nd degree
neighbors scale as $N^{0.15}$. This means that the scaling of the
cover time scales as $N^{0.85}$, as shown in Figure
\ref{poissonscaling}. Note how well the simulation results match
the analytical expression. This is because nodes can be
approximately sampled in an even fashion by following links.

\begin{figure}[tbp]
\begin{center}
\includegraphics[scale=0.5]{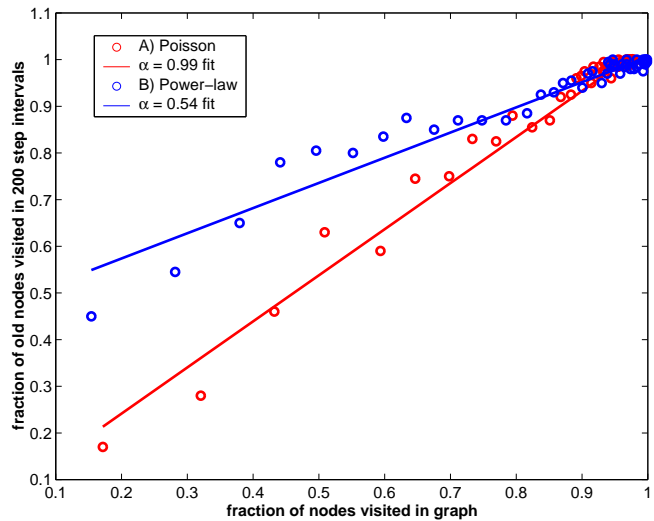}
\end{center}
\caption[Fraction of time nodes are revisited vs. fraction of
nodes visited in the graph]{Fraction of nodes revisited in a given
time interval vs. fraction of nodes visited in the graph for a
search in A) a Poisson graph, and B) a power-law graph.}
\label{fractionrevisit}
\end{figure}

The reason why the cover time for the Poisson graph matches the
analytical prediction and the power-law graph does not is
illustrated in Figure \ref{fractionrevisit}. If links were
approximately evenly distributed among the nodes, then if at one
point in the search 50\% of the graph has already been visited,
one would expect to revisit previously seen nodes about 50\% of
the time. This is indeed the case for the Poisson graph. However,
for the power-law graph, when 50\% of the graph has been visited,
nodes are revisited about 80\% of the time, which implies that the
same high degree nodes are being revisted before new low-degree
ones. It is this bias which accounts for the discrepancy between
the analytic scaling and the simulated results in the power-law
case.

\begin{figure}[tbp]
\begin{center}
\includegraphics[scale=0.5]{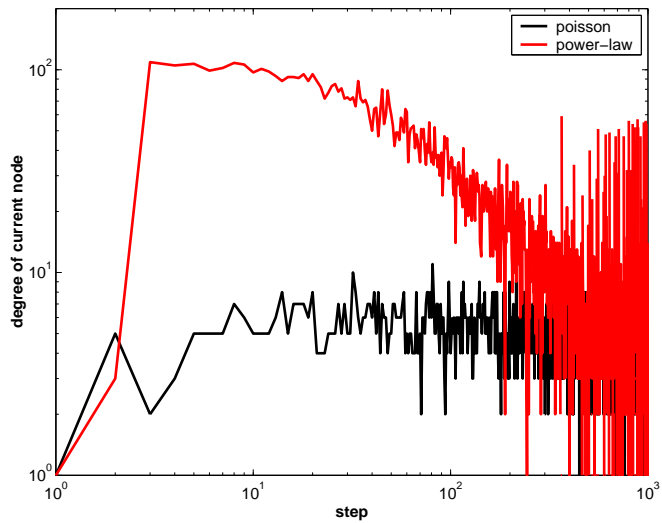}
\end{center}
\caption[Degrees of nodes visited in search]{Degrees of nodes
visited in a single search for power-law and poisson graphs of
10,000 nodes.} \label{DegSequence}
\end{figure}

However, even the simulated $N^{0.35}$ scaling for a random,
minimally self-avoiding strategy on the power-law graph
out-performs the ideal $N^{0.85}$ scaling for the Poisson graph.
It's also important to note that the the high degree node seeking
strategy has a much greater success in the power-law graph because
it relies heavily on the fact that the number of links per node
varies considerably from node to node. To illustrate this point,
we executed the high degree seeking strategy on two graphs,
Poisson and power-law, with the same number of nodes, and the same
exponent $\tau = 2$. In the Poisson graph, the variance in the
number of links was much smaller, making the high degree node
seeking strategy comparatively ineffective as shown in Figure
\ref{DegSequence}.

In the power-law graph we can start from a randomly chosen node.
In this case the starting node has only one link, but two steps
later we find ourselves at a node with the highest degree. From
there, one approximately follows the degree sequence, that is, the
node richest in links, followed by the second richest node, etc.
The strategy has allowed us to scan the maximum number of nodes in
the minimum number of steps. In comparison, the maximum degree
node of the exponential graph is 11, and it is reached only on the
81st step. Even though the two graphs have a comparable number of
nodes and edges, the exponential graph does not lend itself to
quick search.

\section{Gnutella}
Gnutella is a peer-to-peer filesharing system which treats all
client nodes as functionally equivalent and lacks a central server
which can store file location information. This is advantageous
because it presents no central point of failure. The obvious
disadvantage is that the location of files is unknown. When a user
wants to download a file, she sends a query to all the nodes
within a neighborhood of size $ttl$, the time to live assigned to
the query. Every node passes on the query to all of its neighbors
and decrements the $ttl$ by one. In this way, all nodes within a
given radius of the requesting node will be queried for the file,
and those who have matching files will send back positive answers.

This broadcast method will find the target file quickly, given
that it is located within a radius of $ttl$.  However,
broadcasting is extremely costly in terms of bandwidth. Every node
must process queries of all the nodes within a given $ttl$ radius.
In essence, if one wants to query a constant fraction of the
network, say 50\%, as the network grows, each node and network
edge will be handling query traffic which is proportional to the
total number of nodes in the network.

Such a search strategy does not scale well. As query traffic
increases linearly with the size of Gnutella graph, nodes become
overloaded as was shown in a recent study by Clip2
\cite{clip200bwbarrier}. 56k modems are unable to handle more
than 20 queries a second, a threshold easily exceeded by a
network of about 1,000 nodes. With the 56k nodes failing, the
network becomes fragmented, allowing users to query only small
section of the network.

\begin{figure}[tbp]
\begin{center}
\includegraphics[scale=0.5]{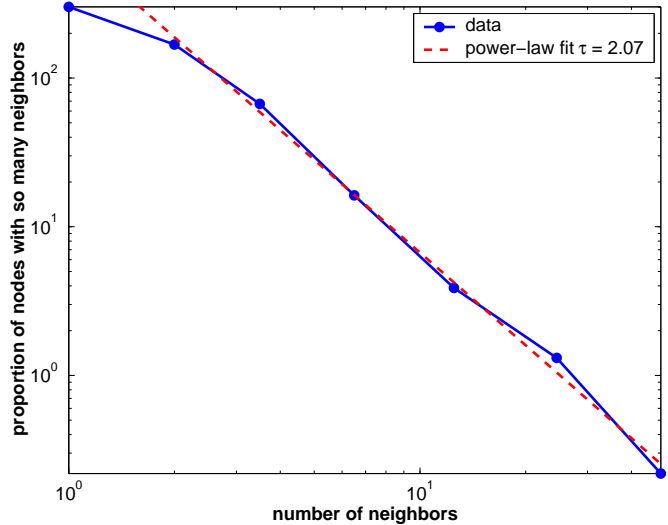}
\end{center}
\caption[Link distribution in the Gnutella network]{Log-log plot
of the link distribution in the Gnutella network. The fitted
exponent $\tau = 2.07$.} \label{gnutellalinkdist}
\end{figure}

The search algorithms described in the previous sections may help
ameliorate this problem. Instead of broadcasting a query to a
large fraction of the network, a query is only passed onto one
node at each step.  The search algorithms are likely to be
effective because the Gnutella network has a power-law
connectivity distribution as shown in Figure
\ref{gnutellalinkdist}.

Typically, a Gnutella client wishing to join the network must find
the IP address of an initial node to connect to. Currently, ad hoc
lists of "good" Gnutella clients exist \cite{clip200bwbarrier}. It
is reasonable to suppose that this ad hoc method of growth would
bias new nodes to connect preferentially to nodes which are
already fairly well-connected, since these nodes are more likely
to be "well-known".  Based on models of graph growth
\cite{barabasi99scaling,adamic99nature} where the "rich get
richer", the power-law connectivity of ad hoc peer-to-peer
networks may be a fairly general topological feature.

By passing the query to every single node in the network, the
Gnutella algorithm fails to take advantage of its connectivity
distribution. To implement our algorithm the Gnutella clients must
be modified to keep lists of the files stored by their first and
second degree neighbors have\footnote{This idea has already been
implemented by Clip2 in a limited way. 56k modem nodes attach to a
high bandwidth Reflector \cite{clip200reflector} node which stores
the filenames of the 56k nodes and handles queries on their
behalf.}. This information must be passed at least once when a new
node joins the network, and it may be necessary to periodically
update the information depending on the typical lifetime of nodes
in the network. Instead of passing the query to every node,
queries are only passed along to the highest degree nodes. The IP
numbers of the nodes already queried are appended to the query,
and they are avoided.

The modified algorithm places an additional cost on every node,
that of keeping track of the filenames of its neighbors' files.
Since network connections saturated by query traffic are a major
weakness in Gnutella, and since computational and storage
resources are likely to remain much less expensive than bandwidth,
such a tradeoff is readily made. However, now instead of every
node having to handle every query, queries are routed only through
high connectivity nodes. Since nodes can select the number of
connections that they allow, high degree nodes are presumably high
bandwidth nodes that can handle the query traffic. The network has
in effect created local directories valid within a two link
radius. It is resilient to attack because of the lack of a central
server. As for power-law networks in general
\cite{barabasi00error}, the network is more resilient than random
graphs to random node failure, but less resilient to attacks on
the high degree nodes.

\begin{figure}[tbp]
\begin{center}
\includegraphics[scale=0.5]{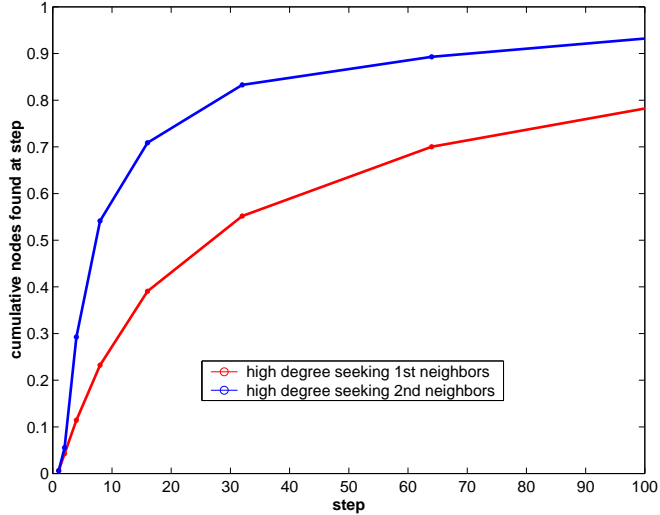}
\end{center}
\caption[Cumulative number of nodes found at each step in the
Gnutella network]{Cumulative number of nodes found at each step in
the Gnutella network.} \label{gnutellapass}
\end{figure}

Figure \ref{gnutellapass} shows the success of the high degree
seeking algorithm on the Gnutella network. We simulated the
search algorithm on a crawl by Clip2 of the actual Gnutella
network of approximately 700 nodes. Assuming that every file is
stored on only one node,  50\% of the files can be found in 8
steps or less. Furthermore, if the file one is seeking is present
on multiple nodes, the search will be even faster.

To summarize, the power-law nature of the Gnutella graph means
that these search algorithms can be effective.  As the number of
nodes increases, the (already small) number of nodes that will
need to be queried increases sub-linearly. As long as the high
degree nodes are able to carry the traffic, the Gnutella network's
performance and scalability may improve by using these search
strategies.

\section{Conclusion}
In this paper we have shown that local search strategies in
power-law graphs have search costs which scale sub-linearly with
the size of the graph, a fact that makes them very appealing when
dealing with large networks. The most favorable scaling was
obtained by using strategies which preferentially utilize the
high connectivity nodes in these power-law networks. We also
established the utility of these strategies for searching on the
Gnutella peer-to-peer network.

It may not be coincidental that several large networks are
structured in a way that naturally facilitates search. Rather, we
find it likely that these networks could have evolved to
facilitate search and information distribution. Networks where
locating and distributing information, without perfect global
information, plays a vital role tend to be power-law with
exponents favorable to local search.

For example, large social networks, such as the AT\&T call graph
and the collaboration graph of film actors, have exponents in the
range ($\tau = 2.1 - 2.3 $) which according to our analysis makes
them especially suitable for searching using our simple, local
algorithms.  Being able to reach remote nodes by following
intermediate links allows communication systems and people to get
to the resources they need and distribute information within these
informal networks.  At the social level, our analysis supports the
hypothesis that highly connected individuals do a great deal to
improve the effectiveness of social networks in terms of access to
relevant resources \cite{gladwell00}.

Furthermore, it has been shown that the Internet backbone has a
power-law distribution with exponent values between 2.15 and 2.2
\cite{faloutsos99topology}, and web page hyperlinks have an
exponent of 2.1 \cite{barabasi99scaling}. While in the Internet
there are other strategies for finding nodes, such as routing
tables and search engines, one observes that our proposed strategy
is partially used in these systems as well. Packets are routed
through highly connected nodes, and users searching for
information on the Web turn to highly connected nodes, such as
directories and search engines, which can bring them to their
desired destinations.

On the other hand, a system such as the power grid of the western 
United States, which does not serve as a message passing network, 
has an exponent $\tau \sim 4$ \cite{barabasi99scaling}. It would 
be fairly difficult to pass messages in such a network without 
knowing the target's location. \\ \\

{\large \textbf{Acknowledgement}}

We would like to thank Clip2 for the use of their Gnutella crawl 
data.

\bibliographystyle{plain}
\bibliography{PLSearch}

\end{document}